\documentclass[%
 reprint,
floatfix,
superscriptaddress,
showpacs,preprintnumbers,
 amsmath,amssymb,
 aps,
pra,
]{revtex4-1}
\usepackage{xcolor}
\usepackage{times}
\usepackage{lineno} 
\usepackage{amssymb}
\usepackage{graphicx}% Include figure files
\usepackage{dcolumn}% Align table columns on decimal point
\usepackage{bm}% bold math
\usepackage{epstopdf}
\usepackage[colorlinks]{hyperref}% add hypertext capabilities
\hypersetup{colorlinks = true,
linkcolor = blue,
filecolor = blue,
urlcolor = blue,
citecolor = blue}
\newcommand*\chem[1]{\ensuremath{\mathrm{#1}}}

\begin{document}
\title{\textbf{Proximity-induced charge density wave in a metallic system}}

\author{Anurag Banerjee}
\affiliation{Department of Physics, Ben-Gurion University of the Negev, Beer-Sheva 84105, Israel}
\affiliation{Institut de Physique Th\'eorique, Universit\'{e} Paris-Saclay, CEA, CNRS, F-91191 Gif-sur-Yvette, France}

\author{Louis Haurie}
\affiliation{Institut de Physique Th\'eorique, Universit\'{e} Paris-Saclay, CEA, CNRS, F-91191 Gif-sur-Yvette, France}

\author{Catherine P\'epin}
\affiliation{Institut de Physique Th\'eorique, Universit\'{e} Paris-Saclay, CEA, CNRS, F-91191 Gif-sur-Yvette, France}

% Please give the surname of the lead author for the running footer
\begin{abstract}
Non-local quasiparticles in correlated quantum materials can exhibit the proximity effect. For instance, in metal superconductor hybrid systems, the leaking of cooper pairs to the metallic region induces superconducting correlations in a standard metal. This paper explores the proximity effects of charge density wave (CDW) on metal using the attractive Hubbard model, which harbors CDW state at half-filling. Our fully self-consistent calculations demonstrate that periodic charge modulations develop in a metal due to the tunneling of finite momentum particle-hole pairs from the CDW region. Upon doping the normal region, the commensurate CDW changes to an incommensurate one by incorporating regular phase shifts. Furthermore, the induced CDW produces a soft gap in the density of states and thus can be detected in tunneling experiments. We discuss our results in light of recent reports of such proximity-induced charge order in different two-dimensional heterostructures. 
\end{abstract}
\maketitle
\section{Introduction}
\label{introduction}
Strongly correlated quantum systems exhibit extremely rich phase diagrams with various broken symmetry phases like superconductivity (SC), charge~\cite{wise_charge-density-wave_2008,chang_direct_2012,Xu_2021} and 
spin density wave~\cite{Cai2013}, Mott insulator~\cite{DagottoRMP,Keimer2015}, among others~\cite{AgterbergPDW,hamidian_detection_2016,PNAS_Nematic1,vestigialNem_PNAS}. Charge density waves (CDW) are often  found close by SC in several transition metal dichalcogenides (TMDs)~\cite{LiuAPL,YanSTM}, cuprates~\cite{wise_charge-density-wave_2008,chang_direct_2012}, and twisted bilayer graphene~\cite{TBG_CDW1,Jiang2019}. Such closeness of quantum many-body phases suggests an intimate connection between CDW and SC despite their vastly differing physical properties~\cite{loret_intimate_2019}.
 
Moreover, rapid advances in Van der Waals engineering of thin layered two-dimensional (2D) materials allow precise control over their electronic structure via tuning of the doping, strain, and thickness~\cite{BoschkerHetero,heteroWang}. These layers can be stacked with other 2D materials to form a heterostructure like a lego. The different layers can harbor multiple broken symmetry phases and topological band structures, which can be merged to provide desirable properties~\cite{Radisavljevic2011}. Thus, contact proximity effects are now becoming a vital tuning knob to drive correlation effects among different layers of 2D materials. Since the electrons cannot abruptly change their nature across the interface between two materials, proximity-induced order develops in the other region until scattering with free electrons kills the phase coherence~\cite{PdGbook}. Atomically clean interfaces in 2D systems reduce such scattering events allowing a long-ranged proximity-induced order to survive. Early studies on SC/metal hybrid systems establish the tunneling of SC pair amplitude in the metals~\cite{MeissnerSC,clarke1968proximity}. Additionally, proximity effects are found for magnetism~\cite{MagneticProximity}, topological insulators~\cite{TopologyProximity}, and quantum Hall systems~\cite{ChiralQH_Finkle}. 

However, the CDW proximity effect of $1T$-\chem{TaS_2} on thin Bismuth has only been recently reported~\cite{BisumuthAPL}. The proximity effect disappears as the thickness of the Bismuth layer increases, showing CDW correlations decay in the three-dimensional limit. Scanning tunneling
microscopy (STM) studies also show proximity-induced CDW from $1T$-\chem{TaS_2} on graphene~\cite{altvater_revealing_2022}. The induced CDW in graphene shows the same `star-of-David' pattern as in the TMD layer. CDW Proximity effects are 
also apparent in similar hybrid materials from the resistivity measurements~\cite{kimCDW2022}. Another study focuses on CDW and SC proximity effects of $1H$-\chem{NbSe_{2}} on few different materials
and found CDW proximity effects on some among those~\cite{Dreher21Nbse2}.
Additionally, in cuprates/magnetites heterostructure, CDW order stabilizes by the inverse proximity effect of ferromagnetism on SC order~\cite{Frano16Cuprates}. 
These experimental signatures prompt a more detailed understanding of CDW proximity effects.

Early theoretical works using semi-classical analysis predict proximity-induced CDW generated by Friedel oscillations~\cite{MesoscopicCDW,rejaeicollective96}. However, self-consistent microscopic calculations for such proximity-induced CDW order are still lacking, unlike superconducting order~\cite{black-induced_2013}. Moreover, the connection of induced CDW with the Friedel oscillations needs to be carefully examined. Furthermore, the effect of different tuning parameters like doping and disorder must be clarified.
The behavior of experimentally relevant physical quantities like the local density of states (LDOS) and spectral functions for the proximity-induced CDW needs to be comprehended. 

This paper focuses on a minimal model to study CDW proximity effects on metal. The attractive Hubbard model, which supports a charge density wave at half-filling~\cite{MoreoPRL,MicnasRMP}, acts as our model for the CDW layer. Similarly, a non-interacting tight-binding model on a square lattice characterizes the metal. We allow the electrons to hop back and forth from the CDW region to the metal region. Strikingly, such a straightforward model can account for the CDW correlations in normal regions. As the hole-doping increases on the metal side, we observe a transition from commensurate induced CDW to incommensurate by incorporating regular phase shifts. Furthermore, the ordering wavevector of the CDW oscillation is distinct from that of the Friedel oscillation close to half-filling. The destruction of the induced CDW on disordering the CDW side also distinguishes between the proximity effect and the Friedel oscillation. We predict the behavior of the local density of states at Fermi energy follows the induced CDW structure, which can be extracted in STM experiments. In the next section, we discuss the details of the model and method used in this study.

\section{Model and method}
We study the effective tight-binding attractive Hubbard model to capture the interplay between the charge density wave (CDW) on a metallic system. The model is given by,
\begin{equation}
\mathcal{H}=-t\sum_{\langle i,j \rangle, \sigma} \left(  c^{\dagger}_{i  \sigma} c_{j \sigma} + H.c. \right) -\sum_{i} U_i \hat{n}_{i,\uparrow} \hat{n}_{i,\downarrow} -\sum_{i,\sigma} \mu_i \hat{n}_{i,\sigma}
\end{equation}
Here $c^{\dagger}_{i,\sigma} (c_{i,\sigma})$ creates (annihilates) electron of spin $\sigma=\uparrow,\downarrow$ at site $i$ in a two-dimensional square lattice. The parameter $t$ denotes the hopping strength between the nearest neighbor site. We fix $t=1$, and all the energy scales are in the units of $t$. $U_i$ is the site-dependent on-site attractive potential, and $\mu_i$ is the site-dependent chemical potential term. The density operator is denoted by $\hat{n}_{i\sigma}=c^{\dagger}_{i,\sigma} c_{i,\sigma}$. The local density is given by  $\rho_i=\sum_\sigma \langle c^{\dagger}_{i,\sigma} c_{i,\sigma} \rangle$, where $\langle ... \rangle$ denotes the expectation value in the ground state since we work at zero temperature, i.e. $T=0$. 

\begin{figure}[h!]
\includegraphics[width=0.475\textwidth]{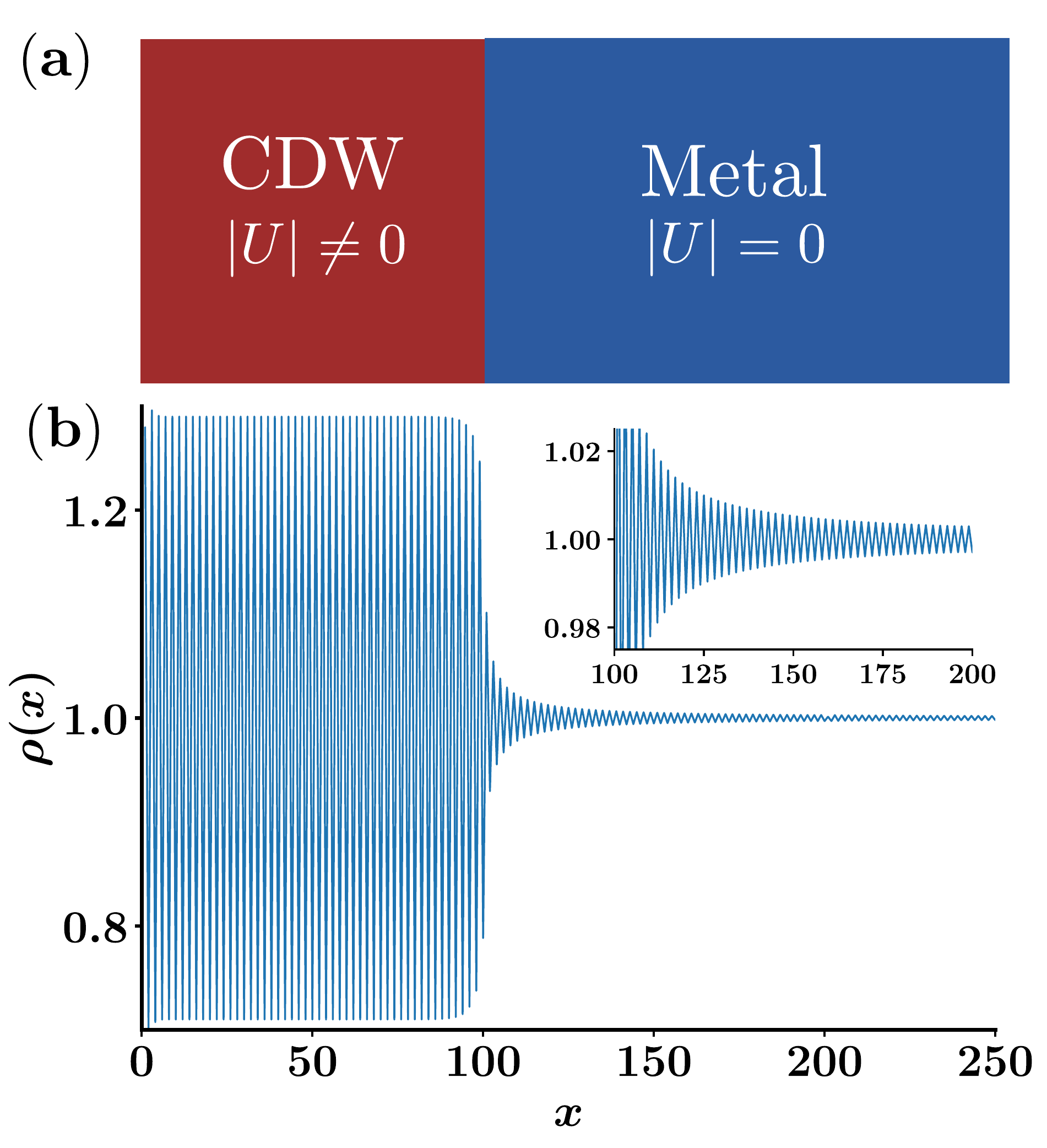}
\caption[0.5\textwidth]{(a) Shows the schematics of the model setup. The left side allows for a charge density wave state due to the attractive interactions at half-filling. The interaction is set to zero on the right side leading to a non-interacting region for $L>100$. (b) Shows the self-consistent local density along the $x$-direction. The region where the interactions are present shows strong density wave oscillations with a modulation wavevector $Q=\pi$. However, CDW modulations survive on the non-interacting side, leading to proximity-induced CDW fluctuations. Inset shows local density in the non-interacting regions. }
\label{fig:fig1}
\end{figure}

To model a CDW-Metal junction, we choose the on-site attractive potential $U_i$ in the following manner,
\begin{align}
U_i =  \begin{cases}
U & \mbox{for   } x_i \leq L_J,\\
0 & \mbox{for }   x_i>L_J,
\end{cases} 
\label{eq:MR_app_11}
\end{align}
where $L_J$ is the length of the interacting side. The schematic of our setup is shown in Fig.~(\ref{fig:fig1}a), where on the left side, correlated phases can generate from the interaction term. In contrast, the right-hand side is just the non-interacting tight-binding model, which is metallic in the absence of any proximity effect. Electrons can hop from the interacting side to the non-interacting side generating proximity effects. For the rest of the analysis, the interaction is fixed to $U=1.5t$. 

Similarly, the average electron density on the left and right terminals is tuned by modifying the chemical potential,
\begin{align}
\mu_i =  \begin{cases}
0 & \mbox{for   } x_i \leq L_J,\\
-\mu & \mbox{for }   x_i>L_J,
\end{cases} 
\label{eq:MR_app_12}
\end{align}
The average density of the electron is given by ${\rho=(1/N)\sum^N_{i=1} \rho_i}$, where $N$ is the number of lattice sites in each terminal. Here we work with the convention such that the half-filling is at $\mu=0$, and the negative chemical potential reduces the average electron density (hole-doped). As seen in Eq.~(\ref{eq:MR_app_12}), we always fix the density of the interacting side at half-filling. However, we study in Sec.~(\ref{subsec:doping}) the effect of hole-doping the non-interacting region and represent the average density of the non-interacting region by $\rho_N$.

The attractive interactions between the electrons allows for two-independent orders at half-filling -- s-wave superconductivity and charge density waves. Therefore we make a mean field decomposition of the interaction term in the Cooper and Hartree channels.
The inhomogeneous mean field decomposition of Hartree shift, where $\rho_i=\sum_{\sigma}\langle c^{\dagger}_{i\sigma} c_{i\sigma} \rangle$ is a site-dependent parameter along with local superconducting pairing  $\Delta_i=-U_i \langle c_{i\downarrow} c_{i\uparrow} \rangle$. To find a self-consistent CDW pattern, we provide the initial guess of the local density to be modulating around the average density $\rho$, given as
\begin{equation} 
\rho_i=\rho+\chi_i \cos(\mathbf{Q}.\mathbf{r}).
\end{equation}
Here, $\chi_i$ is the local CDW amplitude and ordering wavevector $\mathbf{Q}=(\pi,\pi)$.  

To focus on the CDW proximity effect, we suppress the superconducting order. The SC proximity effect is well known in similar calculations, although studies for the CDW states are still lacking~\cite{black-induced_2013,ZhuPRB_Proximity,Proximity_Marsiglio}. We can suppress the SC order by setting pairing to $\Delta_i=0$ on all the sites. After performing the Hartree-Fock decomposition of the inhomogeneous attractive Hubbard model, we obtain, 
\begin{equation}
\mathcal{H}^{\rm CDW}= -t\sum_{\langle i,j \rangle, \sigma} \left(  c^{\dagger}_{i  \sigma} c_{j \sigma} + H.c. \right) - \sum_{i, \sigma} \left(\mu_i+\frac{U_i}{2} \rho_i \right) n_{i \sigma}
\label{mfd59}
\end{equation} 
The Hartree term gives rise to the charge density wave for the attractive Hubbard model at half-filling. Therefore we always fix the left terminal at half-filling to produce a CDW state. Finally, we solve for local density until self-consistency is achieved.

An independent calculation is performed with a homogeneous Hartree shift to compare the above mean-field decomposition with no CDW on the left terminal. The Hamiltonian is given by
\begin{equation}
\mathcal{H}^{\rm Hom}= -t\sum_{\langle i,j \rangle, \sigma} \left(  c^{\dagger}_{i  \sigma} c_{j \sigma} + H.c. \right) - \sum_{i, \sigma} \left(\mu_i+\frac{U_i}{2} \rho \right) n_{i \sigma}
\label{mfd61}
\end{equation} 
Here we have replaced the local density with the average density, prohibiting any charge ordering on the interacting side. Such mean-field decomposition distinguishes between the CDW proximity effect and the regular Friedel oscillation due to the electron density mismatch at the two terminals.

In Fig.~(\ref{fig:fig1}a) presents the schematics of our setup.  We have open boundary conditions along the $x$-direction, whereas we used periodic boundary conditions in the $y$-directions. Due to the translation invariance in the $y$-direction, one can use the standard repeated zone scheme to block diagonalize the Hamiltonian. Note that a $(\pi,\pi)$ CDW order breaks the sublattice symmetry of the square lattice, the density wave pattern is periodic after two lattice sites in both directions. Hence, the supercell consists of two consecutive rows of length $L$. We solve for supercell of size $2\times L$ and use the repeated zone scheme following Ref.~\cite{black-induced_2013,ZhuPRB_Proximity,Proximity_Marsiglio}. We performed our calculations on a $400\times 400$ system. We fix the $L_J=100$ and study the proximity effects on the metallic regions for all $x>100$.

\section{Results}
\label{sec:Res}
In Fig.~(\ref{fig:fig1}b), we plot the self-consistent local density profile where both sides are precisely at the half-filling. Since the system supports the charge order on the left side, it shows a robust staggering pattern for $x<L_J$. Interestingly, the CDW modulations impede the non-interacting region. A significant charge oscillation remains at the same wavevector in the normal region for $x>L_J$ as shown in the inset of Fig.~(\ref{fig:fig1}b). Note that the system also exhibits an inverse proximity effect of the metal in reducing the CDW amplitude near the interface. We present our results far from the boundary at $x=400$.

\subsection{Effect of doping the normal region}
\label{subsec:doping}
\begin{figure}[h!]
\includegraphics[width=0.475\textwidth]{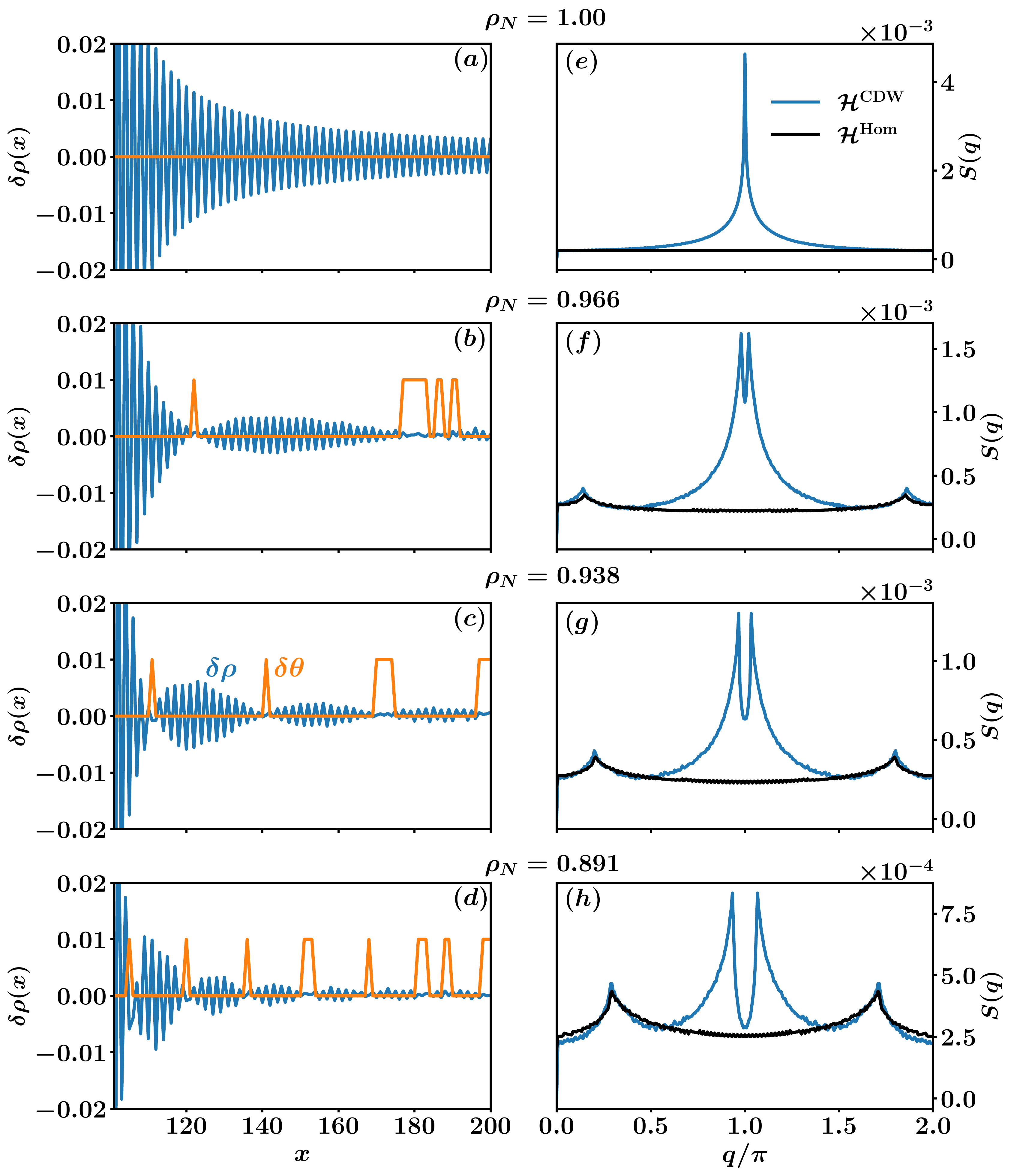}
\caption[0.5\textwidth]{Left panels show the evolution of the proximity-induced local density oscillation with changing electron density $\rho_N$ on the non-interacting side.
(a) $\rho_N=1.0$ shows long-ranged but decaying charge modulations in the non-interacting terminal. (b) For $\rho_N = 0.966$, the local density generates domains of CDW modulations as it suffers regular phase shifts. The orange traces indicate the position of phase shifts.
(c) For $\rho_N = 0.938$, as the non-interacting side is hole-doped further, the coherent domains of CDW modulations become shorter.
(d) For $\rho_N = 0.891$, the phase shits becomes more frequent.
Right panels show the Fourier transform of the density fluctuations identifying the primary ordering wave-vectors of the density modulations.
(e) The $S(q)$ shows the peak at $q=\pi$ as expected. The black traces show the same from an independent calculation with a homogeneous Hartree shift. (f) For $\rho_N=0.966$, the ordering wave-vector splits from $\pi$.
The additional peak at low-$q$ appears due to Friedel oscillation. (g) For $\rho_N=0.938$, the splitting of the ordering peak increases. (h) Same for $\rho_N=0.891$.}
\label{fig:fig2}
\end{figure}
Next, we study the evolution of local density as the average density $\rho_N$ is tuned in the normal region.~\footnote{We cannot tune the average density in the left terminal as the CDW order is only stabilized at half-filling.} Such a study can reveal whether the CDW proximity effect is a special feature of the particle-hole symmetric point exactly at half-filling. The tuning $\rho_N$ only on the normal regions leads to a density mismatch at the interface. Such mismatch leads to the Friedel oscillation along with the CDW proximity effects. We show the density fluctuation $\delta \rho(x)=\rho(x)-\rho$, on the left panels of Fig.~(\ref{fig:fig2}). Furthermore, we track any sign change in the regular density pattern to identify the phase shift of the oscillation and indicate it using the orange traces. On the right panel is the Fourier transform of $\delta \rho$, denoted by $S(q)$. 

Fig.~(\ref{fig:fig2}a) shows regular CDW oscillation with decaying amplitude when $\rho_N=1$. Additionally, there is also no phase shift at this electron density. The Fourier transform of the density fluctuations $S(q)$ shows in Fig.(\ref{fig:fig2}e) a sharp peak at the ordering wavevector $q=\pi$. However, since there is no average density mismatch between the two terminals, the $S(q)$ is flat for $\mathcal{H}^{\rm Hom}$, i.e., for a system with no CDW. This indicates that the charge modulations in the normal region are due to the CDW proximity effect and not regular Friedel oscillations.

As the density of the normal region is tuned away from half-filling, the situation changes significantly. Fig.~(\ref{fig:fig2}b-d) reveals domains of coherent charge modulations with an occasional phase shift. The $S(q)$ peak also displays in Fig.~(\ref{fig:fig2}f-h) a splitting around the ordering wavevector. The amount of splitting is inversely proportional to the mean distance between the successive phase-shift. Furthermore, Friedel oscillations are generated along with the CDW proximity effect due to the density mismatch between the two terminals. The black traces in Fig.~(\ref{fig:fig2}f-h) show the $S(q)$ peak for a system with no charge order in the left terminal. This reveals that the weaker peaks in Fig.~(\ref{fig:fig2}f-h) are due to Friedel oscillations.

As the hole-doping of the non-interacting side increases, the domains of coherent modulations reduce in size. Consequently, the splitting of the peak increases. The split peak from the remnant ordering wavevector almost merges with the Friedel oscillation peak for $\rho_N \sim 0.7$ for the parameters studied in this paper. Hence, the CDW proximity effect, if any, is not distinguishable from the standard Friedel oscillation for such low electron density of the normal region.

\subsection{Local density of states}
\label{subsec:LDOS}
\begin{figure}[ht]
{\includegraphics[width=8.5cm,keepaspectratio]{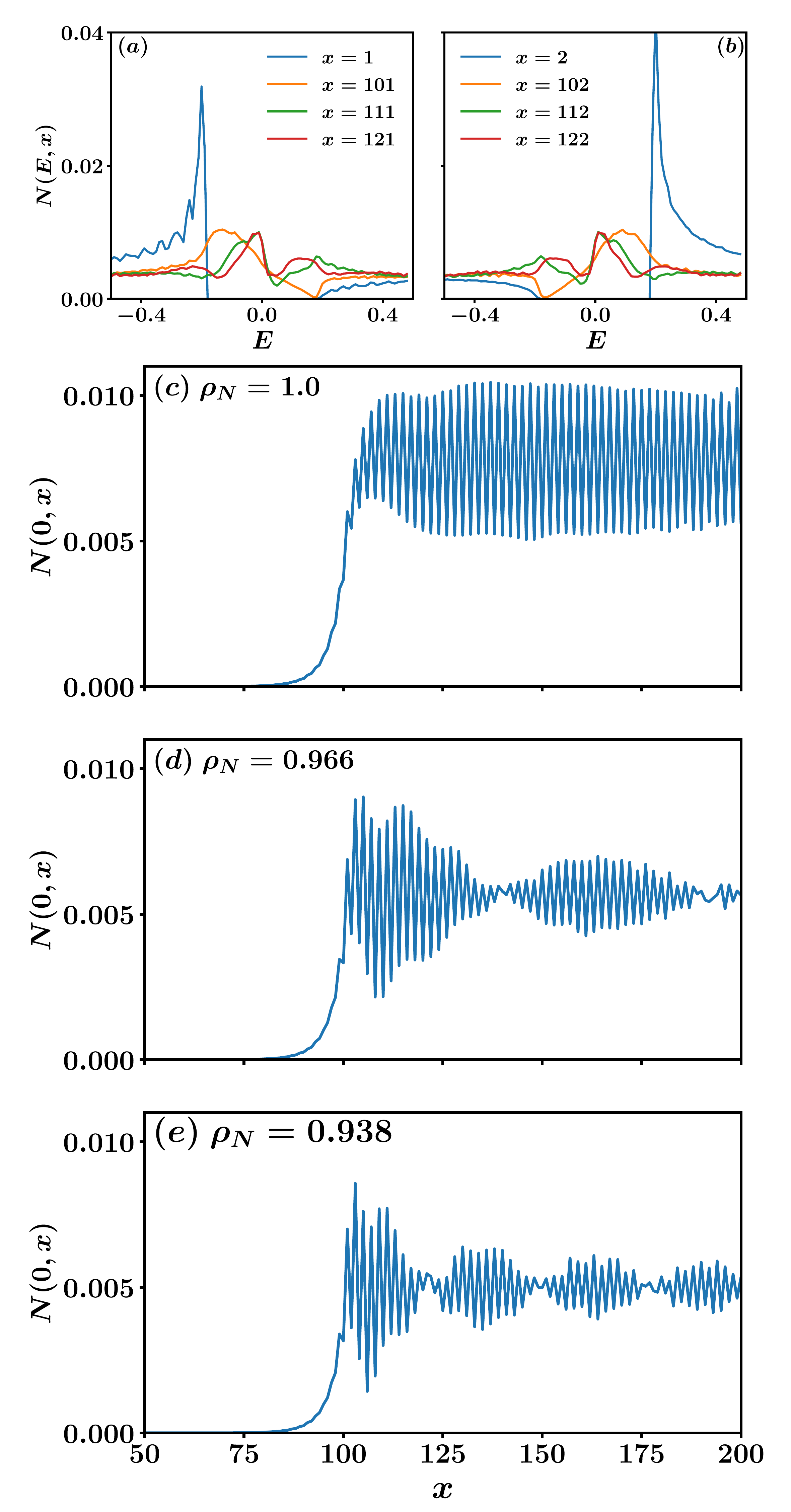}}
\caption{(a-b) Presents the local density of states at location $x$. (a) for A-sublattice and (b) for B-sublattice. For $x<100$, the system is in the CDW phase and consequently gapped. For $x>100$, the LDOS produces in-gap states, and exhibits a dip. (c-e) shows $N(E=0,x)$ for different electronic density of the non-interacting terminal. The spatially resolved LDOS for the interacting region at $E=0$ vanishes in the CDW region but shows density oscillation in the normal region. LDOS at zero energy oscillates spatially with the same wavevector $Q=\pi$. The phase-shift of the $\rho(x)$ observed in Fig.~(\ref{fig:fig2}) can also be identified in $N(E=0,x)$.   
}
\label{fig:fig3}
\end{figure}
We study the local density of states (LDOS), routinely measured in STM experiments. The LDOS is given by
\begin{align}
    N(E,x)=\sum_n \vert \phi_n(x) \vert^2 \delta(E-\epsilon_n),
\end{align}
where $\epsilon_n$ and $\phi_n$ are the eigvenvalues and eigenvectors of the self-consistent mean-field Hamiltonian $\mathcal{H}^{\rm CDW}$.
We quote the energy such that the Fermi energy is always at $E=0$. First, we focus on the LDOS at a particular spatial location $x$ (fixing $y=1$) for $\rho_N=1.0$ in Fig.(\ref{fig:fig3}a,b). As expected, the CDW terminal for $x<100$ has a hard gap. However, the gap edge peak oscillates from positive to negative energies due to the density oscillation in the two sublattices. Such oscillation indicates that at A(B)-sublattice has more occupied (unoccupied) states below(above) the Fermi energy as shown in Fig.(\ref{fig:fig3}a)(b). This makes LDOS a probe of the local electron density studied in the previous section.

Moreover, the gap of LDOS starts to fill up as one approaches the boundary of the CDW region at $x=100$. A soft dip at the gap edge is observed precisely at the interface. Far away from the interface, the magnitude of the dip reduces, and the location of the dip approaches $E=0$. The LDOS of the normal region is unlike a typical metal and is significantly altered due to the induced CDW.

The presence of the in-gap states in the normal region with a gapped spectrum in the CDW region makes it an experimentally accessible tool to detect proximity-induced spatial modulations. First, we dwell on the LDOS at Fermi energy as a function of $x$ in Fig.~(\ref{fig:fig3}c-e). For $\rho_N=1$, the LDOS at $E=0$ modulates with the same wavevector and has no phase shifts as observed in Sec.~(\ref{subsec:doping}). Moreover, as we move away from half-filling, the phase shift can also be identified in the LDOS at zero energy in Fig.~(\ref{fig:fig3}d,e).    

\subsection{Effect of disorder}
\label{sunsec:disorder}
\begin{figure}[ht]
\includegraphics[width=8.5cm,keepaspectratio]{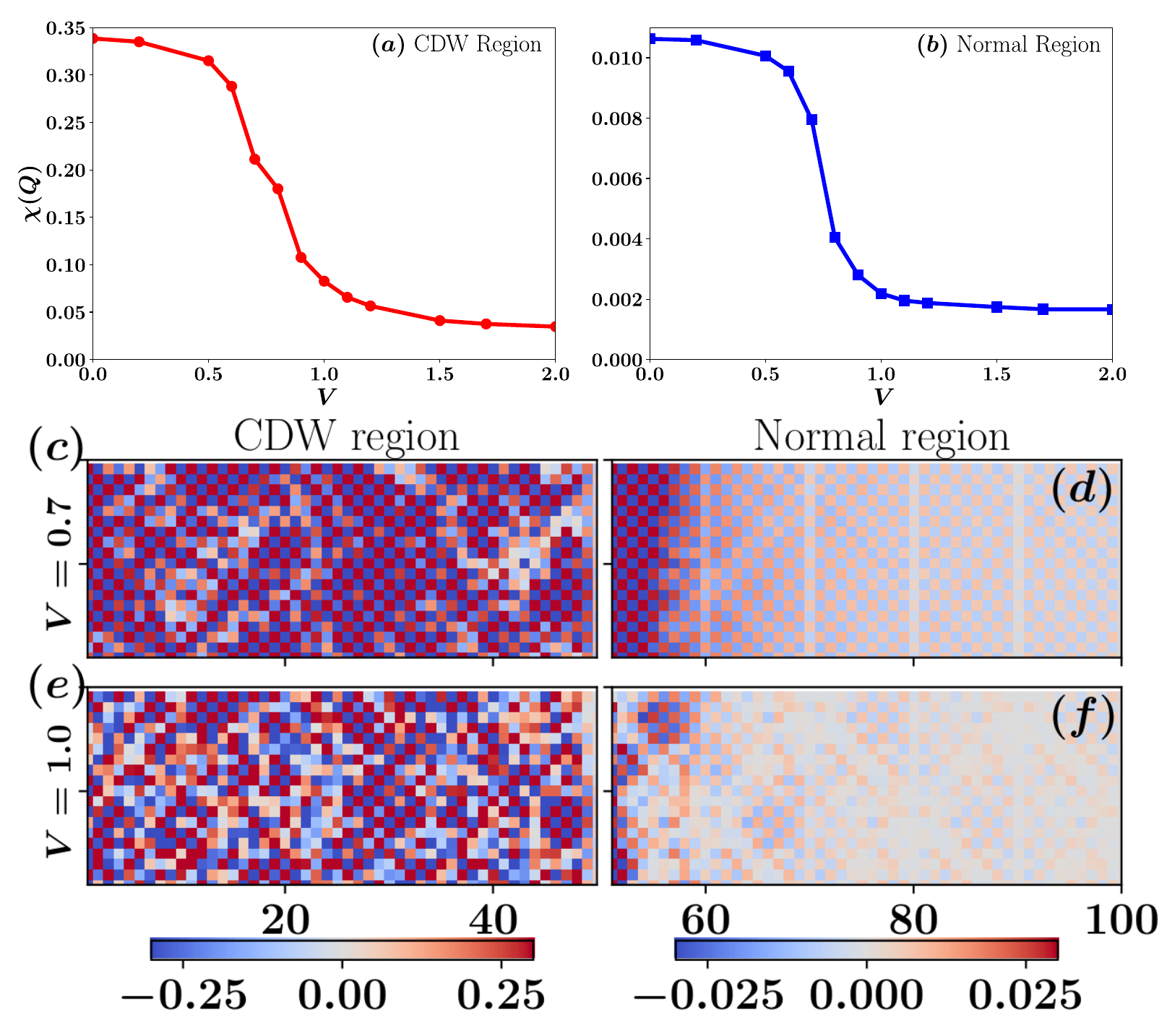}
\caption[0.5\textwidth]{(a) The CDW amplitude reduces with increasing disorder strength for the interacting side. (b) The strength of the proximity-induced CDW amplitude closely follows the demise of the CDW in the interacting region. (c) Shows the density profile $\delta \rho(x,y)$ for $V=0.7$ in the interacting region. (d) Displays the same in the normal region. Even when the density profile for $x<50$ is inhomogeneous due to impurities, the normal region shows uniform proximity-induced charge modulations. However, when the disorder strength restricts the charge modulations in phase-shifted puddles, as shown in (e), only weak patches of proximity-induced charge modulations appear on the normal side shown in (f).
}
\label{fig:fig4}
\end{figure}
Additionally, we perturb the charge ordering by applying disorder in the CDW region and study the proximity of disordered CDW on clean metal. The disorder is modeled by
\begin{align}
    \mathcal{H}_V=\sum_{i,\sigma} V_i \hat{n}_{i\sigma},
\end{align}
where $V_i$ are chosen from a uniform random distribution such that $V_i=[-V/2,V/2]$ with $V$ representing the disorder strength. Since the charge order breaks the translational symmetry which makes it essential to solve the problem in real space only. Hence we work with a smaller system $20\times100$ with $L_J=50$.

Past studies have shown the charge order gets weakened by impurities, as it forms phase-shifted puddles of charge modulations~\cite{HuscroftScalettar,Anurag0}. Fig.~(\ref{fig:fig4}a) confirms that the picture as the mean amplitude of the CDW reduces rapidly as one increases the disorder strength. The $\chi(Q)$ is peak of the $S(q_x,q_y)$ averaged over $20$ independent random configurations. The proximity-induced CDW in the normal region also weakens around the same disorder strength.

Next, we focus on the spatial features of the disordered CDW proximity effect. For $V=0.7$, disorder creates an inhomogeneous modulation pattern in the interacting region. However, it still induces a uniform charge order in the normal region (Fig.~(\ref{fig:fig4}d). Thus, the proximity effect persists if an overall CDW ordering is retained nearby. However, the spatial organization changes drastically when the average ordering collapses for $V=1.0$. Fig.~(\ref{fig:fig4}e) exhibits small phase-shifted puddles of charge modulations in the interacting regions. The proximity-induced charge order is only observed in small patches as the density remains uniform in most samples. Furthermore, $\delta \rho$ only shows mild oscillations due to impurity-induced Friedel oscillations in the nearby region.
\section{Summary and conclusions}
\label{sec:disc}
This paper uses the attractive Hubbard model to study the charge density wave proximity effect on metal. Charge density modulations induce in the normal regions due to the tunneling of finite momentum particle-hole pairs from the interacting to the metallic regions. Therefore, the minimal ingredient to capture the contact proximity effect is the tunneling of quasiparticles from the ordered to the metallic region and vice versa. Although our model assumes a clean interface tunneling at the CDW metal junction, our results are expected to hold for moderately disordered tunneling. Consequently, advancements in fabricating atomically smooth surfaces with the capability to stack them without contamination demonstrate the proximity effects of CDW on graphene from the $1T$-\chem{TaS_2}~\cite{altvater_revealing_2022}.

Furthermore, the induced CDW pattern suffers regular phase shifts upon doping the normal region, leading to the ordering wavevector splitting. Such splitting of the ordering wavevector indicates an incommensuration of the induced charge order. Local density of states can capture such phase shifts in CDW as shown in Fig.~(\ref{fig:fig3}d,e). Past studies~\cite{MesarosPNAS} show a relation between the phase shift and the splitting of the ordering wavevector. The ordering wavevector shifts if the number of periods the commensurate CDW should accommodate over the whole system modifies due to the phase fluctuations. The doping of the normal regions in Fig.~(\ref{fig:fig2}) reduces the number of complete oscillations with the same wavevector. Hence, it suffers phase-shift and splitting of ordering wavevector. Our results predict that the experiments on the graphene/$1T$-\chem{TaS_2} hetero-structures by tuning the electron density in the graphene layer should reveal such qualitative signatures. Furthermore, our study also predicts a CDW proximity effect even if the charge pattern in CDW layer is inhomogeneous due to spatial disorder.

Moreover, in Appendix.~(\ref{Sec:CDWSCProx}), we demonstrate the combined proximity effects when the SC and CDW state coexists at half-filling. 
A particle-hole transformation of the down-electron operators transforms the attractive interactions to repulsion at half-filling~\cite{HuscroftScalettar}.
Such transformation generates a mapping from the coexisting CDW and s-wave superconductivity to the different components of the anti-ferromagnetic
state. Therefore, by such mapping, an antiferromagnetic proximity effect can develop in the repulsive Hubbard model connected to metal.

Although our study is motivated by the recent observations of CDW proximity effects in the TMDs~\cite{altvater_revealing_2022,kimCDW2022,Dreher21Nbse2}, our model is not fine-tuned to capture the physics of such materials. Detailed modeling of interactions, lattice structure, and hoppings should be performed for TMDs in future. Another interesting future direction is to look for proximity-induced charge order in strongly correlated systems~\cite{Anurag1, Choubey_2017}. Also, the effect of thermal fluctuations on the induced CDW needs to be explored in the future.

\section{Acknowledgement}
The authors thank Yvan Sidis for valuable discussions. The numerical calculations were performed on the IPhT cluster Kanta and partially on the BGU HPC clusters. A.B. acknowledges postdoctoral funding from the Kreitman School of Advanced Graduate Studies and European Research Council (ERC) Grant Agreement No. 951541, ARO (W911NF-20-1-0013).
\appendix
\begin{figure}[h!]
\includegraphics[width=0.475\textwidth]{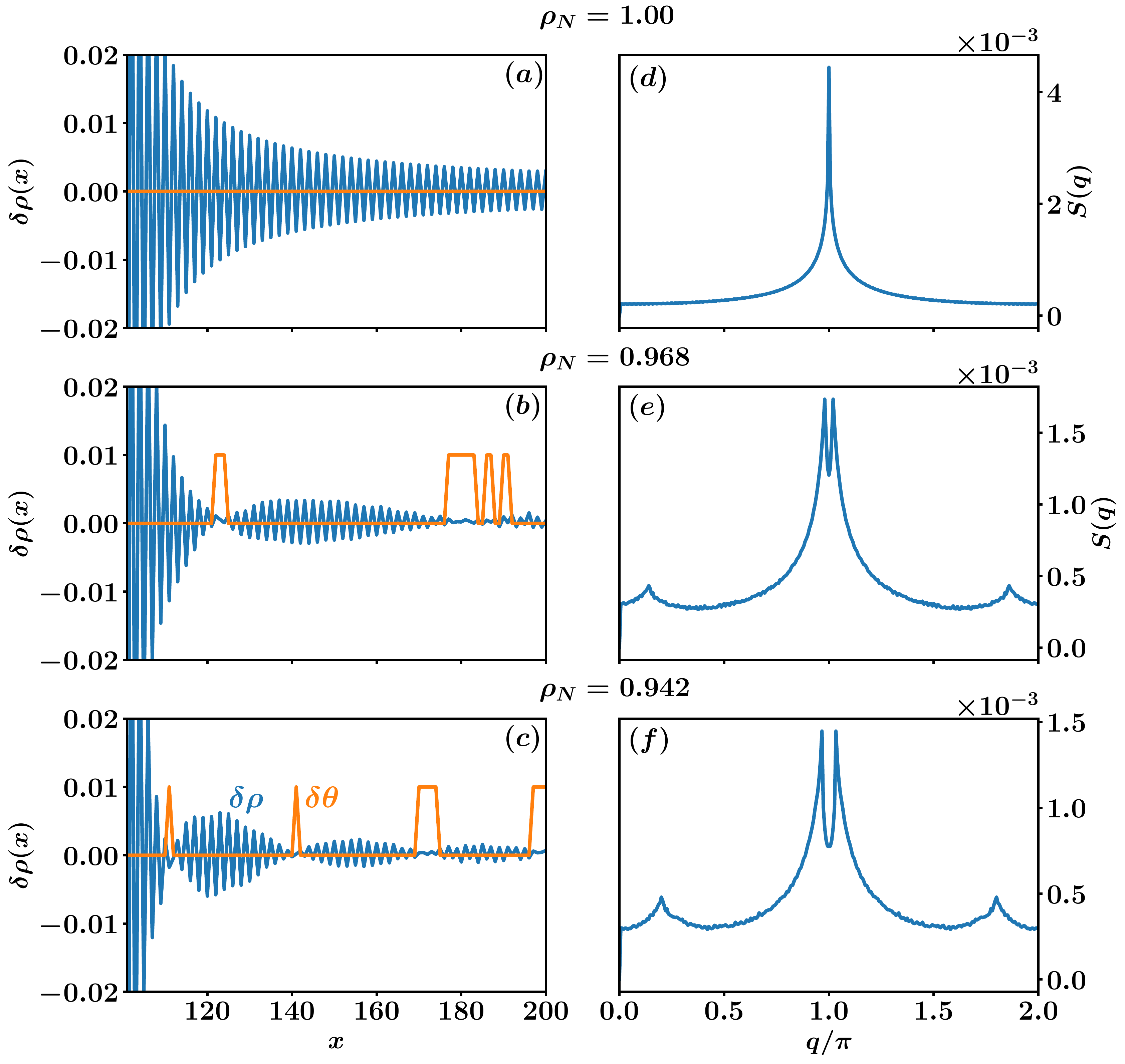}
\caption[0.5\textwidth]{For the nearest neighbor repulsion model, the left panels show the proximity-induced local density oscillation with changing $\rho_N$.
(a) $\rho_N=1.0$ shows long-ranged but decaying charge modulations in the non-interacting terminal. (b) For $\rho_N = 0.968$, the local density suffers regular phase indicated by the orange traces.
(c) For $\rho_N = 0.942$, the coherent domains of CDW modulations further shorten.
The right panels show the Fourier transform of the density fluctuations.
(d) The $S(q)$ shows the peak at $q=\pi$ as expected.  (e) For $\rho_N=0.968$, the ordering wave-vector splits from $\pi$.
The additional peak at low-$q$ appears due to Friedel oscillation. (f) Same for $\rho_N=0.942$.}
\label{fig:figApp}
\end{figure}
\section{Nearest neighbor repulsion model for CDW}
In this appendix, we outline another similar model for charge density wave and study its proximity effect on metal. We consider the electrons on a square lattice with nearest neighbor repulsion. The Hamiltonian is given by
\begin{align}
    \mathcal{H}=-t\sum_{\langle i,j \rangle, \sigma} \left(  c^{\dagger}_{i  \sigma} c_{j \sigma} + H.c. \right) +\sum_{\langle i,j \rangle, \sigma,\sigma^\prime} W_{ij} \hat{n}_{i,\sigma} \hat{n}_{j,\sigma^\prime} \nonumber \\-\sum_{i,\sigma} \mu_i \hat{n}_{i,\sigma}
\end{align}
where $W_{ij}$ is the nearest neighbor repulsion between sites $i$ and $j$. We allow uniform repulsive interaction $W$ in the left terminal while setting it to zero on the right terminal. Thus it is given by
\begin{align}
W_{i,j} =  \begin{cases}
W & \mbox{for   } x_i \leq L_J,\\
0 & \mbox{for }   x_i>L_J,
\end{cases} 
\label{eq:MR_app_21}
\end{align}
The nearest neighbor term generates a $Q=(\pi,\pi)$ charge density wave near half-filling~\cite{Anurag0}. We use $W=0.5t$ and $L_J=100$ to generate the CDW order. We use the form of the chemical potential as presented in Eq.~(\ref{eq:MR_app_12}). We perform an inhomogeneous mean-field decomposition of the $W$-term in the Hartree channel $\rho_i=\sum_{\sigma}\langle c^{\dagger}_{i\sigma} c_{i\sigma} \rangle$ and Fock channel $\Gamma_{ij}=\langle c^\dagger_{i\sigma} c_{j\sigma} \rangle$.
The Fock term modifies the hopping amplitude in the CDW region $\tilde{t} = t+ W\Gamma_{ij} $. We self-consistently calculate the local density and Fock amplitude. 

Similar to the attractive Hubbard model in the previous section, we also observe a proximity-induced charge order for this model. The decaying yet long-ranged charge order is observed when the normal region is half-filled in Fig.~(\ref{fig:figApp}a). This leads to a sharp peak at the ordering wavevector at $q=\pi$ as shown in Fig.~(\ref{fig:figApp}e). Note that the hopping amplitude differs due to a finite Fock term in the left terminal compared to the metallic regions. However, such a lattice mismatch does not introduce phase shifts in the proximity-induced CDW pattern.
However, as we dope the metallic region away from half-filling, the density modulations form short-ranged domains while suffering regular phase shifts as presented in Fig.~(\ref{fig:figApp}b) and Fig.~(\ref{fig:figApp}c). Like the attractive Hubbard model, the ordering wavevector splits as we dope the normal region away from half-filling.

\section{Proximity effect of coexisting CDW and superconducting orders}
\label{Sec:CDWSCProx}
\begin{figure}[ht]
{\includegraphics[width=8.5cm,keepaspectratio]{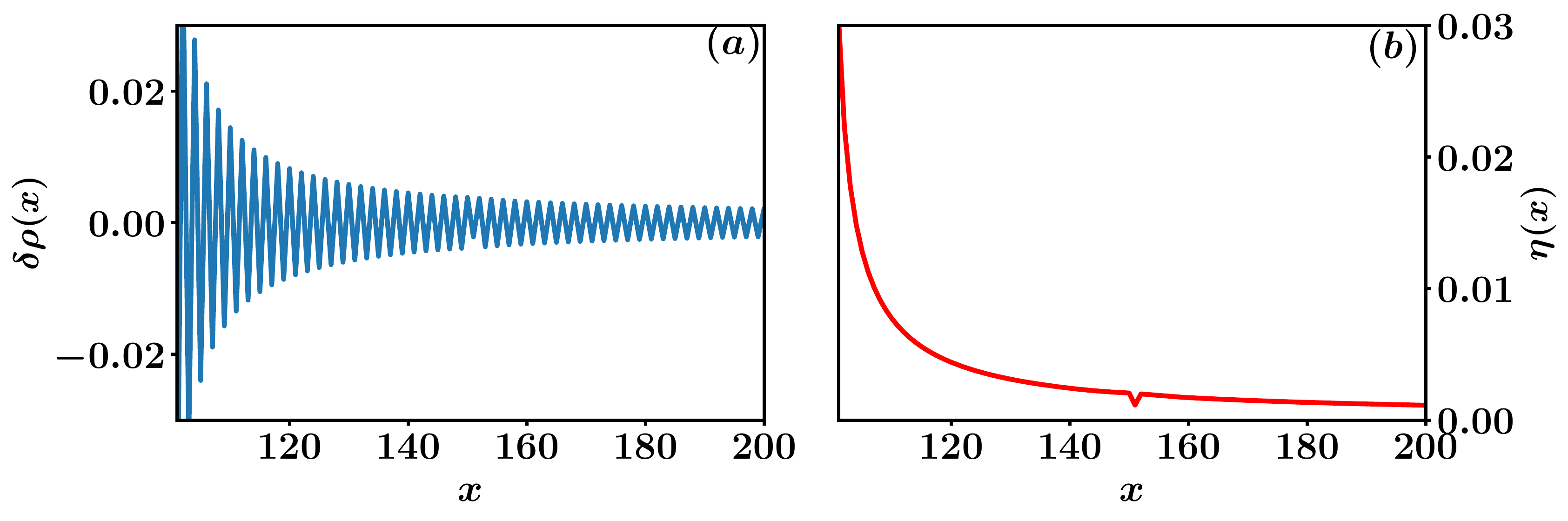}}
\caption{(a) Show the evolution of the proximity-induced local density oscillation in the metallic region when the superconducting order is allowed for the attractive Hubbard model. (b) Displays the proximity-induced Cooper pairing in the metallic region due to the SC order in the left terminal. Thus the CDW proximity effect survives along with the SC proximity effect. We have fixed $U=1.5t$ for a  $300\times 300$ square lattice, and both the terminals are at half-filling with $L_J=100$.
}
\label{fig:figApp2}
\end{figure}
In this appendix, we allow for the superconducting and charge density wave orders in the attractive Hubbard model and test for the proximity effect in the metallic regions. The inhomogeneous mean field decomposition of Hartree shift, where $\rho_i=\sum_{\sigma}\langle c^{\dagger}_{i\sigma} c_{i\sigma} \rangle$ generates the CDW order along with local superconducting pairing  $\Delta_i=-U_i \langle c_{i\downarrow} c_{i\uparrow} \rangle$ in the left terminal.  
After performing the mean-field decomposition in the Hartree and Bogoliubov channels, the mean-field Hamiltonian read
\begin{align}
\mathcal{H}^{\rm CDW}_{\rm SC}= -t\sum_{\langle i,j \rangle, \sigma} &\left(  c^{\dagger}_{i  \sigma} c_{j \sigma} + H.c. \right) - \sum_{i, \sigma} \left(\mu_i+\frac{U_i}{2} \rho_i \right) n_{i \sigma} \nonumber \\&+\sum_{i} \left( \Delta_{i} c^\dagger_{i\uparrow} c^\dagger_{i\downarrow} + H.c. \right) 
\label{mfd58}
\end{align} 
We solve for $\Delta_i$ and $\rho_i$ self-consistently such that the left terminal has coexisting CDW and SC orders. After achieving self-consistency, we check the proximity-induced Cooper pairing by studying  $\eta_i=\langle c_{i\downarrow} c_{i\uparrow} \rangle$ in the metallic region. We solve by setting the interaction $U=1.5t$ for a  $300\times 300$ square lattice when both the terminals are at half-filling with $L_J=100$. 

Due to the interplay of the two orders, the amplitude of both CDW and SC order decreases in the interacting terminals~\cite{MoreoPRL,MicnasRMP}. 
We plot the density-fluctuation $\delta \rho(x)$ in Fig.(\ref{fig:figApp2}a), and it survives in the metallic regions. However, the proximity induced CDW order is weaker 
due to the interplay with the SC order. We also show the proximity-induced Cooper pairing in the metallic region in Fig.~(\ref{fig:figApp2}b). Thus SC proximity effect coexists with the CDW proximity effects in a metal.
\bibliography{Cuprates.bib}
\end{document}